\documentclass[journal]{IEEEtran}
\usepackage{amsmath,amsfonts}
\usepackage{algorithmic}
\usepackage{algorithm}
\usepackage{array}
\usepackage[caption=false,font=footnotesize]{subfig}
\usepackage{textcomp}
\usepackage{stfloats}
\usepackage{url}
\usepackage{verbatim}
\usepackage{graphicx}
\usepackage{cite}
\usepackage[hidelinks]{hyperref}
\usepackage{xcolor}
\usepackage{multirow}
\usepackage{pgfplots}
\usepackage{booktabs}
\usepgfplotslibrary{groupplots,fillbetween}
\pgfplotsset{compat=1.18}
\definecolor{brightpink}{rgb}{1.0, 0.0, 0.5}

\definecolor{forestmydarkgreen(web)}{rgb}{0.13, 0.55, 0.13}

\definecolor{skyblue}{RGB}{51,153,255}

\definecolor{darkgreen}{rgb}{0.13, 0.55, 0.13}

\begin{document}

\title{Degree-Corrected Joint Matrix Factorization for Multilayer Community Detection}

\author{Alexandra Dache, Manon Rustin, Arnaud Vandaele, Nicolas Gillis \thanks{
The authors are with the Department of Mathematics and Operational Research,
University of Mons, Mons, Belgium
(e-mail: firstname.lastname@umons.ac.be).
}}



\maketitle

\begin{abstract}
Multilayer networks allow the modeling of interactions between the same entities across different contexts, such as temporal observations, varying settings, or interactions of different types. The goal of community detection in multilayer networks is to identify groups of nodes exhibiting similar connectivity patterns, which may vary across layers. We propose a method based on a joint nonnegative symmetric matrix trifactorization for community detection in multilayer networks, where each graph is approximated by a nonnegative symmetric matrix trifactorization.  Our approach enforces constraints on the factor matrices so that communities are disjoint and shared across layers, while allowing each layer to have its own connectivity patterns and node degrees. This flexibility enables the model to capture both local and global structural variations across layers. We also develop an algorithm to efficiently solve this problem. We evaluate multilayer community detection methods using the multilayer degree-corrected stochastic block model (MDCBM), a flexible framework for generating realistic multilayer graphs with heterogeneous degrees and varying connectivity patterns. Experiments show that our method reliably detects communities across diverse regimes, whereas existing state-of-the-art approaches are often limited by restrictive structural assumptions.
\end{abstract}


\section{Introduction}
When interaction data are observed across different contexts such as temporal observations, varying environments, or multiple types of relationships, multilayer networks provide a natural way to model this information~\cite{kivela2014multilayer}. A multilayer network, also called a multiplex network in the literature, consists of multiple graphs defined over the same set of nodes.  Each graph represents a layer associated with a specific context or interaction type. Here, we focus on the case without inter-layer connections, although multilayer networks can sometimes refer to more general settings. Such data arise in many real-world applications~\cite{Multilayer_survey}, including social networks where individuals interact across multiple platforms~\cite{hamed2024comprehensive}, transportation systems where locations are connected through different modes of transport~\cite{multilayer_public_transport}, and biological systems~\cite{GOSAK2018118}.

Community detection in multilayer networks aims to identify groups of nodes that exhibit similar connectivity patterns across layers. In this work, we adopt a more general definition of communities, as in stochastic block models, not limited to assortative (densely connected) structures, but based on shared connectivity patterns. Under this definition, nodes belong to the same community if they have similar patterns of interactions with all communities, rather than simply having many connections within their own community.

The goal is to capitalize on the fact that multilayer networks often provide multiple observations of an underlying structure. In many cases, a common community structure is assumed across layers, where each layer provides a different and possibly complementary view of the same structure. Even when this assumption does not strictly hold, it can still be useful to compute a consensus partition that summarizes the network across layers. However, the structural properties of networks may vary significantly across layers, with differences in connectivity patterns and node degree distributions, making community detection more challenging. Methods therefore need to be flexible enough to account for these differences across layers in order to recover shared communities.  This problem is closely related to multi-view clustering~\cite{MC_survey} where data from multiple views are jointly clustered to obtain a clustering solution that uncovers the common latent structure shared across views. Community detection in multilayer networks can be seen as a particular case of multi-view clustering where each view corresponds to a graph. 

Several strategies have been proposed for multilayer community detection~\cite{Multilayer_survey,multiplex_survey}. Flattening approaches reduce the multilayer network to a single graph  to use single-layer community detection algorithms~\cite{chen2017multilayer,lei2023bias}. However, these approaches can lose layer-specific structure and produce communities driven by edges spread in different layers rather than present within any individual layer. Aggregation methods first detect communities independently in each layer and then try to find a consensus partition~\cite{berlingerio2013abacus,tagarelli2017ensemble}. They assume that each layer can provide a reliable community structure on its own. They may be sensitive to layers with weak or noisy community structures. Direct methods, instead, detect the communities directly on the multilayer network and partition the nodes on all the layers simultaneously~\cite{OLMF,Joint_spectral,S2-jNMF}. 

Among direct approaches, nonnegative matrix factorization (NMF) has been successfully used for clustering and community detection due to its ability to provide interpretable low-rank representations of nonnegative data~\cite{he2021survey,wang2011community}. This approach has naturally been extended to community detection in multilayer networks\cite{S2-jNMF,NMF_aggreg} and to multi-view clustering~\cite{khan2022multi}. NMF aims to approximate nonnegative data through the product of lower-dimensional nonnegative matrices. In the multilayer setting, the adjacency matrix of each layer is decomposed into a nonnegative factorization, while some factors are shared across layers and others are allowed to vary from one layer to another. Additional constraints can also be imposed to improve the interpretability of the factors. The objective is to accurately reconstruct the adjacency matrices while ensuring that the shared factors capture the common community structure across layers and the layer-specific factors capture information specific to each layer.

In this article, we propose mFROST, a community detection method for multilayer networks based on joint nonnegative symmetric matrix trifactorization. To the best of our knowledge, mFROST is the first NMF method that jointly estimates shared community memberships, together with node sociability parameters and interaction matrices specific to each layer. This formulation allows both node sociability and interactions between communities to vary across layers. The model provides an interpretable representation of multilayer networks by separating community memberships, node sociability, and community interactions and is closely related to the multilayer degree-corrected stochastic block model (MDCBM)~\cite{Joint_spectral}, which is a multilayer version of the degree-corrected stochastic block model~\cite{SBM_degreecorection}. We develop an efficient algorithm to solve the factorization problem. We evaluate our method on synthetic multilayer networks generated using the MDCBM. Results show that the proposed approach effectively detects communities under diverse network conditions and remains robust compared to existing methods with stronger structural assumptions. We further validate the method on real-world networks with diverse sizes, degree distributions, proportions of isolated nodes, and community size imbalances, to assess its performance across varied structural conditions.

The rest of this paper is organized as follows. Section~\ref{sec: NMF background} provides background on NMF, with a focus on its application to community detection. Section~\ref{sec: our method} introduces the proposed NMF model and the optimization algorithm. For simplicity, we postpone the review of previous works related to our model to Section~\ref{sec: related work}. Section~\ref{sec: experiments} presents the experimental results. Section~\ref{sec:conclusion} concludes the paper.

\section{NMF Background}
\label{sec: NMF background}
Nonnegative Matrix Factorization decomposes a nonnegative matrix $X\in\mathbb{R}^{n\times m}_+ $ as the product of two low-rank nonnegative matrices $X\approx WH$ with $W\in \mathbb{R}^{n\times r}_+$ and $H\in \mathbb{R}^{r\times m}_+$, where $r\ll\min(m,n)$~\cite{lee1999learning}. Using the Frobenius norm to measure the reconstruction error, the NMF problem is:
\[\min_{W\geq 0,\;H\geq 0} \|X-WH\|_F^2.\] 
NMF provides a linear dimensionality reduction of the data, where $W$ contains the basis vectors of a latent $r$-dimensional space and $H$ contains the corresponding coefficients for each data point. The nonnegativity constraints provide an interpretable representation, as the coefficients in $H$ quantify the contribution of each latent component to the data points. Therefore, a clustering assignment can be directly obtained by selecting the component with the largest coefficient for each data point. An explicit hard clustering can be obtained by imposing the orthogonality constraint $H H^\top = I_r$, where $I_r$ is the identity matrix of dimension $r$, so that the rows of $H$ are orthonormal. Together with the nonnegativity constraint, this implies that each column of $H$ contains at most one nonzero entry. This constrained formulation, known as orthogonal nonnegative matrix factorization (ONMF), is closely related to k-means clustering~\cite{POMPILI201415}.

For symmetric matrices, such as the adjacency matrices of undirected graphs, symmetry preservation leads to a better representation of the data. Symmetric nonnegative matrix factorization (SymNMF) addresses this by imposing a symmetric factorization of the form: $X\approx ZZ^\top$, where $Z(i,:)^\top\in \mathbb{R}_+^r$ indicates the membership of node $i$ in each of the $r$ communities. SymNMF has been successfully applied to graph clustering~\cite{SymNMF} and to community detection~\cite{wang2011community}. However, the factorization $X\approx ZZ^\top$ works best when there are no interactions between the communities and can only produce positive semidefinite matrices $X$. To overcome these limitations, symmetric nonnegative matrix trifactorization (TrisymNMF) introduces an additional interaction matrix $\theta$, such that $X\approx Z\theta Z^\top$,  where $\theta\in\mathbb{R_+}^{r\times r}$ captures interactions between the communities. For example, \cite{BNMTF} applies this formulation to community detection. In this paper, we use the orthogonal symmetric nonnegative matrix trifactorization (OtrisymNMF), an extension of TrisymNMF that imposes the additional orthogonality constraint $Z^\top Z=I_r$ to obtain disjoint communities.

\section{Our model}

\label{sec: our method}
\begin{figure}[!t]
\centering
\hfill
\subfloat[Assortative]{%
    \includegraphics[width=0.155\textwidth]{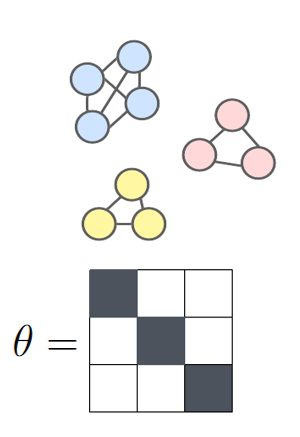}%
    \label{fig:assortative}%
}
\hfill
\subfloat[Disassortative]{%
    \includegraphics[width=0.155\textwidth]{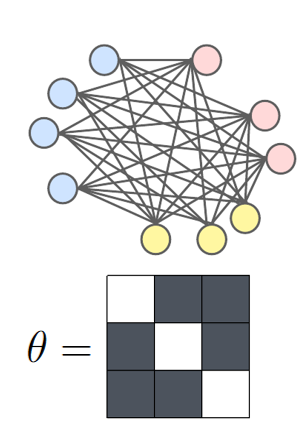}%
    \label{fig:disassortative}%
}
\hfill
\subfloat[Chain-like]{%
    \includegraphics[width=0.155\textwidth]{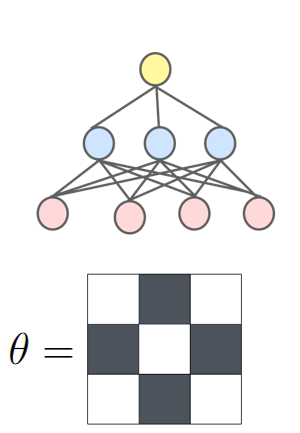}%
    \label{fig:chainlike}%
}
\caption{
Examples of connectivity patterns that can be represented by OtrisymNMF with 3 communities, illustrated with the matrix $\theta$, where entries with high values are shown in black. 
\label{fig:structures}
}
\end{figure}

We consider a multilayer graph with $n$ nodes and $L$ layers, represented by the adjacency matrices $A_l, l=1,\dots,L$, and aim to partition the nodes into $r$ communities. We propose to approximate each graph by a nonnegative matrix trifactorization:
\[A_l\approx Z_l\theta_l Z_l^{\top} \quad\text{s.t.} \quad Z_l^\top Z_l=I_r,Z_l\geq 0,\theta_l\geq0,\] 
where $Z_l\in \mathbb{R}_+^{n\times r}$ and $\theta_l\in\mathbb{R}^{r\times r}_+$. The nonnegativity constraints on $Z_l$, together with the orthogonality constraints on the columns of $Z_l$, imply that each row of $Z_l$ contains at most one nonzero entry. These constraints allow for highly interpretable parameters. Specifically, each edge between two nodes $i$ and $j$ is approximated as: \[A_l(i,j)\approx Z_l(i,k)\theta_l(k,m) Z_l(j,m),\] 
where $k$ and $m$ are the column indices of the nonzero entries in rows $i$ and $j$ of $Z_l$. The matrix $Z_l$ identifies the community memberships: if $Z_l(i, k)> 0$,
node $i$ belongs to community $k$, and the value of $Z_l(i,k)$ quantifies the sociability of node $i$. Indeed, a larger value of $Z_l(i,k)$ leads to a larger expected connectivity of node $i$ with other nodes. The matrix $\theta_l$ captures the connectivity pattern, that is, the interactions between communities, where $\theta_l(k, m)$ quantifies the strength of the connection between communities $k$ and $m$. The matrix $\theta_l$ is not constrained to be diagonal, so communities can interact. This model is known as orthogonal symmetric nonnegative matrix trifactorization (OtrisymNMF) and was originally introduced in~\cite{OtriNMF}. OtrisymNMF can model a wide range of connectivity patterns and combinations of these; see Fig.~\ref{fig:structures} for an illustration.

For the multilayer setting, we must determine which factors are shared across layers and which are layer-dependent. The goal is to find communities common to all layers, which implies that all matrices 
$Z_l$ have the same community assignments: each row has at most one nonzero entry, and its position is the same across layers.  However, the values of $Z_l$ are allowed to vary across layers, as a node may exhibit different sociability levels in different layers. Accordingly, we decompose $Z_l=D_lV$, where $D_l$ is an $n\times n$ nonnegative diagonal matrix containing the values of $Z_l$, and $V$ is a binary $n \times r$ matrix with exactly one entry equal to $1$ per row, assigning each node to a community. We have that $Z_l(i,k)=D_l(i,i)$ if $V(i,k) =1$, meaning that node $i$ is assigned to community $k$. Note that this model handles isolated nodes. If a node $i$ is isolated in layer $l$, we have that $D_l(i,i)= 0$. The matrix $\theta_l$ is layer-dependent, since the communities can interact differently across layers.  For each layer $l$, the adjacency matrix is then approximated as \begin{equation*}
    A_l\approx D_lV\theta_l(D_lV)^\top.
    \
\end{equation*}
This formulation is closely related to the multilayer degree-corrected stochastic block model (MDCBM)~\cite{Joint_spectral}, in which the expected adjacency matrix is given by  $\mathbb{E}[A_l]=D_lV\theta_l(D_lV)^\top$. Further details on the MDCBM are provided in Section~\ref{sec: related work}. Finally, to detect communities across layers with adjacency matrices $A_l$, we propose to solve:\begin{equation}
\begin{aligned}
&\min_{D_l\geq0,\,V\in\{0,1\}^{n\times r},\,\theta_l\geq0} \quad 
 \sum_l \|A_l - Z_l \theta_l Z_l^\top\|_F^2 \\
& \quad \quad \quad \text{s.t.} \quad 
Z_l^\top Z_l = I_r,\quad Z_l = D_l V.
\end{aligned}
\label{Joint OtrisymNMF}
\end{equation}
The objective is the reconstruction error, defined as the sum of squared Frobenius reconstruction errors across layers. The shared assignment matrix $V$ identifies the communities and determines the possible positions of nonzero entries in each matrix $Z_l$.

\subsection{mFROST: Our Algorithm for~\eqref{Joint OtrisymNMF}} 

We propose an alternating optimization algorithm for problem~\eqref{Joint OtrisymNMF}. Our method, called mFROST, is based on the single-layer OtrisymNMF algorithm introduced in~\cite{dache2026matrixfactorizationframeworkcommunity}. It employs an alternating optimization procedure commonly used in NMF methods. The matrices $Z_l$ and $\theta_l$ are first initialized. Then, at each iteration, the matrices $Z_l$ are optimized while fixing matrices $\theta_l$, then, matrices $\theta_l$ are optimized while fixing matrices $Z_l$. These two steps are repeated until the difference in relative error 
\begin{equation}
    \sqrt{\frac{\sum_l \|A_l - Z_l \theta_l Z_l^\top\|^2_F}{\sum_l\|A_l\|_F^2}},
    \label{eq: relative error}
\end{equation}
between two consecutive iterations is smaller than $\delta$, or a maximum number of iterations is reached. Our main algorithmic contribution concerns the update of $Z_l$: we introduce a joint block coordinate descent scheme that enforces $Z_l=D_lV$ by simultaneously updating the same row in all $Z_l$ matrices, selecting a shared community assignment while optimizing their layer-specific values.

In the numerical experiments, we set $\mathrm{MaxIter}=50$ and $\delta = 10^{-5}$. Across all tested cases, including large real-world datasets, convergence was achieved in fewer than $20$ iterations (see Table~\ref{tab:runtime}). Because the optimization problem is non-convex and mFROST relies on an iterative procedure, its solution may depend on the initialization. This is a common limitation of many iterative optimization methods. Different initializations may therefore lead to solutions with different reconstruction errors. We thus run mFROST with multiple restarts, each with a different initialization, and retain the solution with the smallest reconstruction error. Rather than relying on purely random initialization, we propose a more informative strategy to obtain better starting points. Note that this strategy still involves randomness. In Section~\ref{Section: scalability}, we show that our initialization strategy requires fewer iterations to meet the stopping criterion and consistently reaches better solutions than random initialization. The proposed optimization procedure is summarized in Algorithm~\ref{alg:mFROST}.

\paragraph{Update of $\theta_l$} The update of each matrix $\theta_l$ depends only on the adjacency matrix $A_l$ and the matrix $Z_l$ of its corresponding layer. For fixed $Z_l$, the unconstrained problem is:
\[\min_{\theta_l} \|A_l-Z_l\theta_l Z_l^\top\|_F^2 ,\] and the optimality conditions are given by \[2Z_l^\top(Z_l\theta_l Z_l^\top-A_l)Z_l=0.\]
Since $Z_l^\top Z_l=I_r$, the optimal solution for $\theta_l$ has a closed form:
\begin{equation}
    \theta_l^*=Z_l^\top A_l Z_l
    \label{eq: theta}
\end{equation} with $\theta_l^*$ symmetric and nonnegative. 

\paragraph{Update of $Z_l$}The update of the matrices $Z_l$ is more challenging since they are constrained to share the same community assignments: Each row of $Z_l$ contains at most one nonzero entry, and this entry must appear at the same position across all layers. To handle this constraint, we propose a block coordinate descent strategy that updates the $i$-th row of all matrices $Z_l$ simultaneously while keeping the other rows fixed. For node $i$, the update consists of finding the position $k\in\{1, \dots ,r\}$ of the possible nonzero entry shared by all layers (i.e., the community assigned to node $i$) and the corresponding values $Z_l(i,k)$ for each layer. To do so, we test each possible community assignment $k=1,\ldots,r$ for the node $i$. For a candidate community $k$, the coefficient $Z_l(i,k)$ is obtained by solving, independently for each layer l, the one-dimensional problem:
\begin{align}
    &\min_{Z_l(i,k)\geq0} \|A_l-Z_l\theta_l Z_l^\top\|_F^2 \nonumber\\
    = &\min_{Z_l(i,k)\geq0} (A_l(i,i)- Z_l(i,k)\theta_l(k,k)Z_l(i,k))^2 \nonumber\\
    &+ 2\sum_{j\neq i} (A_l(i,j)-Z_l(i,k)\theta_l(k,:)Z_l(j,:)^\top)^2 + \text{const.}
    \label{eq: subproblemZ}
\end{align}
Solving~\eqref{eq: subproblemZ} corresponds to minimizing a fourth-order polynomial of the form:
\[
az^4 + bz^2 + cz,
\]
where the coefficients are given by:
\begin{itemize}
    \item $a = \theta_l(k,k)^2$,
    \item $b = 2\left( \sum_{j \neq i} (Z_l(j,:) \theta_l(:,k))^2 - \theta_l(k,k) A_l(i,i) \right)$,
    \item $c = -4 \sum_{j \neq i} A_l(i,j)\, Z_l(j,:) \theta_l(:,k)$.
\end{itemize}
This subproblem is solved by computing the extrema of the fourth-order polynomial using Cardano’s method and selecting the optimal nonnegative value for $Z_l(i,k)$. 
After evaluating all $r$ candidate communities, node $i$ is assigned to the community that yields the largest decrease in reconstruction error, and the corresponding rows $Z_l(i,:)$ are updated for all layers. Once all rows have been updated, the columns of each matrix $Z_l$ are normalized, which does not affect the objective since any column scaling of $Z_l$ can be absorbed into $\theta_l$ in the product $Z_l\theta_lZ_l^{\top}$. Then, the optimal $\theta_l^*$ is computed in~\eqref{eq: theta} for each normalized $Z_l$. 
\paragraph{Convergence}
For each layer, the closed-form update of $\theta_l$ minimizes the objective for fixed $Z_l$. The update of the matrices $Z_l$ is performed by block coordinate descent, with the rows updated simultaneously across layers, and the configuration yielding the smallest total reconstruction error is selected. Each update therefore decreases or leaves unchanged the objective value. Since the objective is bounded below by zero, the sequence of objective values converges.

\paragraph{Initialization}

We only need to initialize the matrices $Z_l$, since $\theta_l$ can be computed directly using~\eqref{eq: theta}.  First, a single-layer community detection algorithm is applied independently to each layer, producing $L$ node partitions. For this step, we use the initialization procedure for OtrisymNMF proposed in~\cite{dache2026matrixfactorizationframeworkcommunity} based on smoothed separable NMF and Orthogonal NMF. This procedure relies on a randomized method and may therefore yield different solutions across runs. This method is chosen because it has polynomial-time complexity and does not rely on assumptions about the underlying community structure. Moreover, in the single-layer case, this method exactly recovers $Z$ in the noiseless setting, and comes with theoretical robustness guarantees in the presence of noise~\cite{dache2026matrixfactorizationframeworkcommunity}.  Then we compute a consensus partition from the $L$ partitions via Ultra-Scalable Ensemble Clustering (U-SENC)~\cite{huang2019ultra}, which applies spectral clustering on the bipartite graph connecting nodes and partitions. The resulting consensus partition defines the support of each $Z_l$. For each layer, the nonzero entries of $Z_l$ are initialized proportionally to the node degrees computed from $A_l$, after which the columns of $Z_l$ are normalized.

\paragraph{Complexity}

Our method mFROST is designed to be scalable to large multilayer networks. In our implementation, each matrix $Z_l=D_lV$ is represented by a layer-specific vector $d_l$ of length $n$, storing its values (i.e., the diagonal entries of $D_l$), together with a single vector $v$ of length $n$, which stores the community assignment of each node, i.e., the column index of the possible nonzero entry in each row of $Z_l$. In short, $ d_l(i)=Z_l(i,v(i)).$ With this representation and appropriate pre-computations, each iteration of the algorithm requires $\mathcal{O}(r\sum_l\text{nnz}(A_l))$ operations, where $\text{nnz}(A_l)$ denotes the number of nonzero entries of the matrix $A_l$. The initialization procedure has the same asymptotic complexity. The total memory complexity is $\mathcal{O}\left(Lnr+Lr^2+\sum_l\mathrm{nnz}(A_l)\right)$. mFROST is thus scalable and computationally
cheap, scaling linearly with the product of the number of edges and the
number of communities.

\begin{algorithm}
\caption{mFROST}
\label{alg:mFROST}
\textbf{Input:} Adjacency matrices $\{A_l\}_{l=1}^{L}$, number of nodes $n$,  number of communities $r$.

\textbf{Output:} Community assignments $v$, community membership matrices $\{Z_l\}_{l=1}^{L}$, community interaction matrices $\{\theta_l\}_{l=1}^{L}$

\begin{algorithmic}[1]

\FOR{$t=1$ to \textit{N\_Restarts}}
    \STATE Initialize $\{Z_l\}_{l=1}^{L}$, $v$ using the consensus-based strategy.
    \STATE Compute $\{\theta_l\}_{l=1}^{L}$ using (\ref{eq: theta})
    \STATE $\mathrm{iter}\leftarrow0$
   \WHILE{Decrease in~\eqref{eq: relative error} between iterations $> \delta$ \textbf{and} $\mathrm{iter}  < \mathrm{MaxIter}$}
   \STATE $\mathrm{iter}  \leftarrow \mathrm{iter}  + 1$
        \FOR{$i=1$ to $n$}
        \FOR{$k=1$ to $r$}
       
        \FOR{$l=1$ to $L$}
         \STATE Solve (\ref{eq: subproblemZ}) to obtain the optimal candidate $x_{lik}\geq0$.
        \ENDFOR
        \ENDFOR
        \STATE Select $k^\star$ whose candidate row updates yield the smallest total reconstruction error across all layers.
        \STATE $v(i) \leftarrow k^\star$
        \FOR{$l=1$ to $L$}
            \STATE $Z_l(i,:) \leftarrow 0$
            \STATE $Z_l(i,k^\star) \leftarrow x_{li k^\star}$
        \ENDFOR
        \ENDFOR
        \STATE Normalize columns of $\{Z_l\}_{l=1}^{L}$
        \STATE Compute $\{\theta_l\}_{l=1}^{L}$ using (\ref{eq: theta})
        
    \ENDWHILE
     \STATE Store $v^{(t)}\leftarrow v$, $\{Z_l^{(t)}\}_{l=1}^{L} \leftarrow \{Z_l\}_{l=1}^{L}$, $\{\theta_l^{(t)}\}_{l=1}^{L} \leftarrow \{\theta_l\}_{l=1}^{L}$, and reconstruction error~\eqref{Joint OtrisymNMF}
    
\ENDFOR
\STATE Select the restart $t^*$ with minimum reconstruction error
\STATE \textbf{return} $v^{(t^*)}$, $\{Z_l^{(t^*)}\}_{l=1}^{L}$, $\{\theta_l^{(t^*)}\}_{l=1}^{L}$

\end{algorithmic}
\vspace{2pt}
\textit{Note: $v(i)$, the community assignment of node $i$, defines the position of the possible nonzero entry for all rows $Z_l(i,:)$, $l=1,\dots,L$.}
\end{algorithm}

\subsection{Related Work}
\label{sec: related work}
In this section, we review multilayer community detection methods that aim to identify groups of nodes with similar connectivity patterns across layers. A fundamental framework for modeling such community structure is provided by stochastic block models. Their multilayer extensions are widely used for theoretical analysis and as benchmark  models to evaluate community detection methods.  To recover such community structures from observed multilayer networks, existing approaches can be broadly divided into three categories: spectral methods, probabilistic approaches, and matrix factorization methods.

The SBM was introduced for community detection in single-layer networks. It models a network by assuming that the probability of an edge between two nodes depends only on their community memberships. The SBM can capture a wide range of connectivity patterns, including assortative, disassortative, and core-periphery structures. The degree-corrected stochastic block model (DCBM)\cite{SBM_degreecorection} generalizes the SBM by accounting for variations in node degrees within the same community.   Among its multilayer extensions, the multilayer degree-corrected stochastic block model (MDCBM)~\cite{Joint_spectral} is the model that best matches our setting: it assumes a shared community structure across all the graphs while both the block connectivity
matrices $\theta_l\geq 0$ and the node degree parameters $d_l(i)\geq 0$ can be different for each layer. The MDCBM assumes that edges of the networks are mutually independent, and their expected values (probabilities) are described by $ \mathbb{E}[A_l(i,j)]= d_l(i)\,\theta_l(k,m)\, d_l(j),$ where $k$ and $m$ are the communities of the nodes $i$ and $j$. Using matrix notation, we have:
\begin{equation*}
    \mathbb{E}[A_l]= D_lV\theta_l V^\top D_l^\top,
\end{equation*}
where $D_l$ is a diagonal matrix with the node degree parameters $D_l(i,i)=d_l(i)$ and $V$ is a binary matrix indicating the community memberships. For unweighted graphs, the expected value of an adjacency entry corresponds to the probability that the associated edge exists. Each adjacency entry is therefore modeled as a Bernoulli random variable:
\begin{equation}
     A_l(i,j)=A_l(j,i) \sim \text{Bernoulli} \left( \left(D_lV\theta_l V^\top D_l^\top\right)_{i,j}\right),
     \label{eq: mdcbm}
\end{equation}
for all $i<j$ and layers $l$.

Spectral methods obtain a low-dimensional representation of the nodes from the leading eigenvectors of a graph matrix, typically the adjacency matrix or the Laplacian. A clustering method is then applied to the resulting representations to obtain the community assignments. A straightforward extension of spectral clustering to multilayer networks consists of applying spectral clustering to the average adjacency matrix $\frac{1}{L}\sum_l A_l$~\cite{han2015consistent,bhattacharyya2020consistent}. This flattening approach assumes similar connectivity patterns across layers and may fail under heterogeneous structures~\cite{lei2023bias}. To address this limitation, the bias-adjusted spectral method based on the sum of squared adjacency matrices was proposed in~\cite{lei2023bias}. Another approach aggregates the projections onto the principal subspaces of each graph~\cite{MASE,Joint_spectral}. These methods first compute a spectral embedding for each layer independently and then concatenate the embeddings across layers.  Singular value decomposition (SVD) is subsequently applied to obtain a shared low-dimensional representation, on which a clustering algorithm is applied to infer the community memberships. In the presence of heterogeneous node degrees, Agterberg et al.~\cite{Joint_spectral} introduce an additional normalization step before concatenation to account for node-specific degree variations in each layer.

Probabilistic approaches model multilayer networks through stochastic block models and infer the latent community structure using statistical inference techniques such as expectation-maximization~\cite{EMforMultilayerMixedMembershipSBM}, variational inference~\cite{ali2019latent}, or Bayesian Markov chain Monte Carlo (MCMC) sampling \cite{peixoto2015inferring}.

Among matrix factorization approaches, Paul and Chen~\cite{OLMF} proposed to solve the orthogonal linked matrix factorization (OLMF):
\begin{equation*}
    \min_{Z,\theta_l}\sum_l \|A_l- Z\theta_lZ^\top\|_F^2 \quad \text{s.t.} \quad Z^\top Z=I_r.
\end{equation*}
No nonnegativity constraints are imposed on $Z$ and $\theta_l$, and $Z$ is shared across layers. The matrix $Z$ is expected to capture the common characteristics of the nodes. The communities are estimated by using k-means to cluster the rows of $Z$. In their publicly released implementation, however, $A_l$ is by default replaced with the regularized graph Laplacian: 
\begin{equation*}
    L_{\tau,l} = \mathcal{D}_{\tau,l}^{-1/2} A_l \mathcal{D}_{\tau,l}^{-1/2}, \qquad \mathcal{D}_{\tau,l} = \mathcal{D}_l + \tau I_n,
\end{equation*}
where $\mathcal{D}_l= \operatorname{diag}(A_l\mathbf{1})$ is a diagonal matrix containing the node degrees in layer $l$ and $\tau$ is set to the average node degree.  This regularization was proposed to stabilize spectral clustering under the degree-corrected stochastic block model in the single-layer setting~\cite{regularizedSC}. This normalization implicitly attenuates the influence of high-degree nodes. More recently, \cite{hribar2026solvingsymmetricmultitypeorthogonal} proposed a joint nonnegative symmetric matrix trifactorization with orthogonality constraints, where $Z$ is shared across layers. 
\begin{equation*}
    \min_{Z\geq 0,\theta_l\geq0}\sum_l \|A_l- Z\theta_lZ^\top\|_F^2 \quad \text{s.t.} \quad Z^\top Z=I_r.
\end{equation*}  Since the same matrix $Z$ is used for all layers, the relative sociability of nodes within a community is constrained to remain the same across layers. For example, consider two nodes $i$ and $j$ belonging to the same community. If node $i$ is more sociable than node $j$ in one layer, but less sociable in another layer, this change cannot be represented by a shared $Z$. The layer-specific matrix $\theta_l$ cannot compensate for this discrepancy, since it modifies the connectivity between communities uniformly for all nodes belonging to the same communities. Consequently, this formulation does not explicitly account for layer-specific variations in node sociability. In Section~\ref{sec:ablation}, we compare shared and layer-specific $Z_l$, with and without the regularized graph Laplacian. We show that allowing $Z_l$ to vary across layers is more effective at handling node degree heterogeneity. For this factorization with a shared $Z$, they proposed two algorithms to solve this problem. The first relies on multiplicative updates, while the second uses Adam, a gradient-based optimization method. Such algorithms cannot enforce nonnegativity and orthogonality simultaneously, and iterates are not feasible. 
Gligorijevi\'c et al.~\cite{NMF_aggreg} proposed another strategy based on one of these four NMF models: SymNMF with orthogonality $A_l\approx ZZ^\top$, Projective NMF $A_l\approx Z Z^\top A_l$, and OtrisymNMF $A_l \approx Z\theta_lZ^\top$ and its semi-NMF variant with $\theta$ not constrained to be nonnegative. In the authors’ publicly available implementation, each adjacency matrix is normalized by the node degrees before factorization. The method has two steps. First, the selected NMF is computed for each layer separately to obtain the layer-specific $Z_l$. Second, all layers are factorized collectively with a single common matrix $Z$, while a penalty term with a hyperparameter $\alpha$ keeps $Z$ close to the layer-specific $Z_l$. The algorithm relies on multiplicative updates. Only the variants called CSNMTF and CSsNMTF based on OtrisymNMF allow layer-specific interactions between communities. The symmetric and projective NMF variants impose the same block structure in all layers. In all variants, $Z$ is shared across layers even if keeping close to the layer-specific representation $Z_l$, so the limitation on node sociability described above applies as well. Finally, semi-supervised joint NMF~\cite{S2-jNMF} was specifically developed for detecting assortative community structures. It first creates new adjacency matrices for each layer by integrating prior information from well-connected motifs across all network layers, extracted using a greedy search method. It then jointly decomposes the
matrices into a basis matrix and multiple coefficient
matrices. The communities are discovered based on the
common basis matrix.


\section{Experiments}
\label{sec: experiments}
To evaluate our method for community detection in multilayer networks, we conduct experiments on synthetic networks generated from the MDCBM and multilayer networks based on real-world datasets. We compare mFROST with the baseline methods evaluated in~\cite{Joint_spectral} and an additional NMF-based method, collective symmetric nonnegative matrix trifactorization (\textbf{CSNMTF})~\cite{NMF_aggreg}.

Spectral methods include spectral clustering on the average adjacency matrix (\textbf{Sum A})~\cite{han2015consistent,bhattacharyya2020consistent}, spectral clustering based on the bias-adjusted sum of squared adjacency matrices (\textbf{Bias-adjusted SoS})~\cite{lei2023bias}, Multiple Adjacency Spectral Embedding (\textbf{MASE})~\cite{MASE}, and its degree-corrected version (\textbf{DC-MASE})~\cite{Joint_spectral}. All spectral methods use spherical spectral clustering by normalizing the rows of the embedding before k-means clustering, following the empirical robustness observed in~\cite{Joint_spectral} under severe degree heterogeneity. We further use the Bayesian inference method implemented in \textbf{graph-tool}~\cite{graph-tool}, which relies on a multilayer degree-corrected stochastic block model with multilevel Markov chain Monte Carlo (MCMC) optimization and a fixed number of communities $r$, and the Orthogonal Linked Matrix Factorization method (\textbf{OLMF})~\cite{OLMF}.  We use the code repository accompanying~\cite{Joint_spectral}, which provides the original implementations of baseline methods as well as the synthetic MDCBM benchmark used in their study. Unless otherwise specified, we retain the default parameter settings and experimental protocol of \cite{Joint_spectral} for all baseline methods. For CSNMTF, we adapt the authors’ publicly available CSNMF code. Both the per-layer and collective factorizations use the same number of restarts as mFROST, retaining the solution with the lowest objective value at each stage. 

For mFROST, we set the stopping tolerance to $\delta=10^{-5}$ on the difference in relative error  between consecutive iterations, with a maximum of $50$ iterations as a safeguard. In practice, mFROST consistently converges well before this limit is reached (see Section~\ref{Section: scalability} for empirical convergence statistics). By default, we use 10 restarts per run. For the synthetic experiments, we reduce this to 3 restarts, since results are already averaged over 100 independent network samples, which already accounts for most of the variance due to initialization. All experiments were performed on a laptop with 2.80 GHz Intel® Core™ i7-1165G7 with 16 GB of RAM. All experiments can be reproduced using the code available on GitHub \url{https://github.com/Alexia1305/mFROST}.

\subsection{Synthetic MDCBM}
\begin{figure*}[!t]
    \centering
    \includegraphics[width=0.8\linewidth]{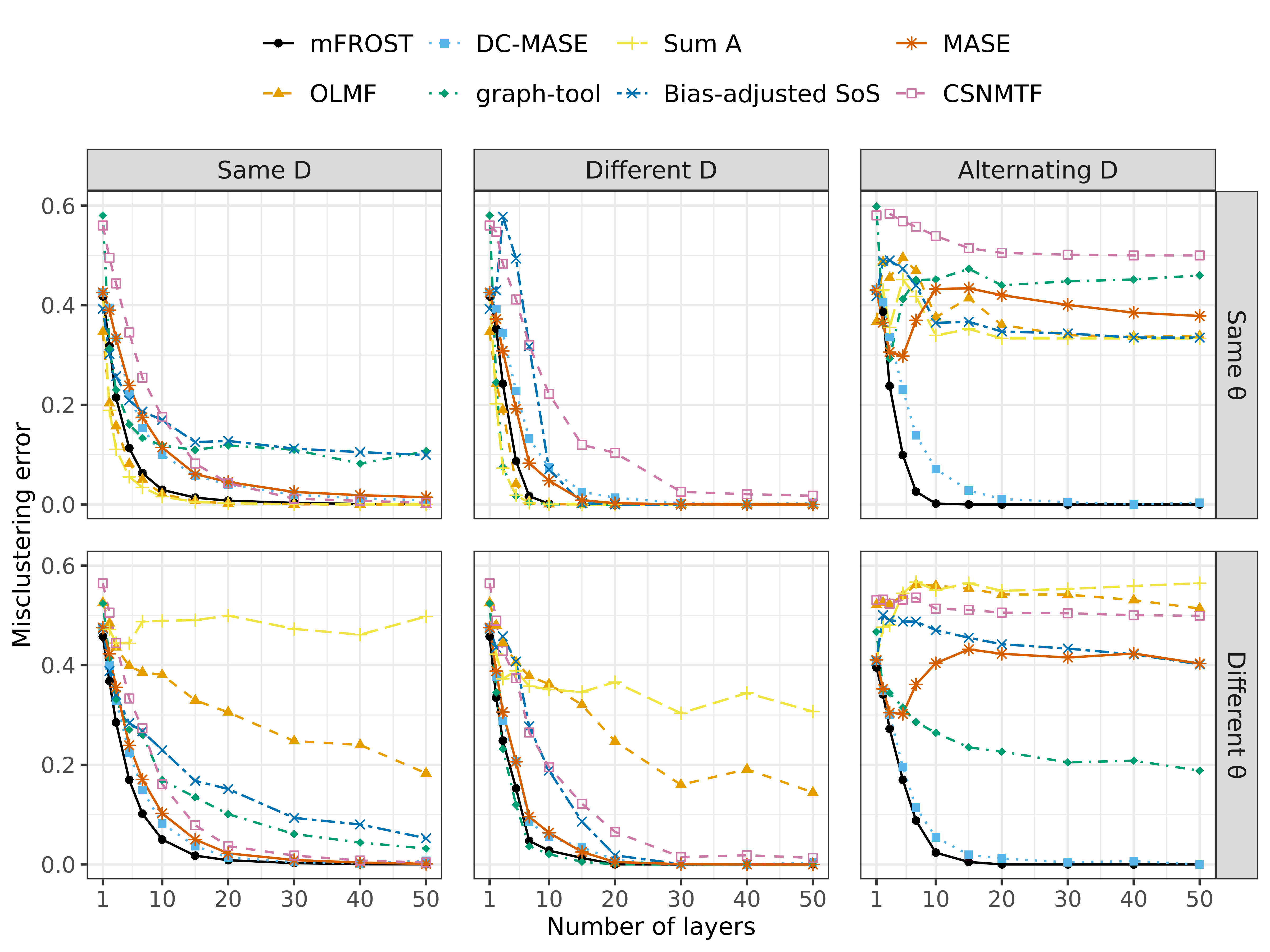}
    \caption{Average misclustering error as a function of the number of layers under different configurations of the connectivity matrices $\theta_l$ and node degree parameters $D_l$.}
    \label{fig:mainexp}
\end{figure*}

We first reproduce the synthetic experiments from \cite{Joint_spectral}, which evaluate the methods for community detection in networks generated from the MDCBM under different parameter setups. All the graphs are generated using the MDCBM~\eqref{eq: mdcbm} with $n=150$ nodes and $r=3$ equal-sized communities. The experiments investigate the effect of the number of layers $L$ on community recovery for different types of heterogeneity. For the connectivity matrices $\theta_l$, we consider two scenarios:
\begin{enumerate}
    \item \textbf{Same connectivity matrices}: identical across layers, with $\theta_l(k,k)=1$ and $\theta_l(k,m)=0.4$ for $m\neq k$.
    \item \textbf{Different connectivity matrices}: generated independently for each layer, with $\theta_l(k,k)=p_l$ and $\theta_l(k,m)=q_l$ for $m\neq k$, where $p_l, q_l \sim \text{Unif}(0,1)$ are drawn independently across layers.
\end{enumerate}
For the sociability level parameters $D_l$, which control node degrees in each layer, we consider three scenarios:
\begin{enumerate}
    \item \textbf{Same degrees}: the degree distribution is heterogeneous across nodes but identical across layers, with $d(1),\dots,d(n) \sim \text{Exp}(1)+0.2$ drawn i.i.d.\ once and shared by all layers.
    \item \textbf{Different degrees}: generated independently for each layer, with $d_l(1),\dots,d_l(n) \sim \text{Exp}(1)+0.2$ drawn i.i.d.\ for each layer $l$.
    \item \textbf{Alternating degrees}: nodes within each community are split into two equal-sized groups. One group has high degree parameters in odd layers and low degree parameters in even layers, and the other follows the opposite pattern. Specifically, $d_l(i)=0.8$ if $l$ and $i$ have the same parity, and $d_l(i)=0.15$ otherwise.
\end{enumerate}
The average degree of the nodes is controlled to remain approximately equal to 10 across all generated networks. For each setting,
the results are averaged over 100 samples. Performance is evaluated using the misclustering error rate, which is the proportion of nodes that are incorrectly clustered.

For CSNMTF, we tested $\alpha \in \{0.001, 0.01, 0.1, 0.5, 1\}$ with $L \in \{5,20\}$ across all six scenarios of the first synthetic experiment. We selected $\alpha = 0.1$, which achieved the lowest misclustering error across these configurations, and fixed it for the remaining experiments.

Figure~\ref{fig:mainexp} shows the results. We observe that mFROST achieves the best performance in scenarios with either different connectivity matrices across layers or alternating degree distributions. When the connectivity matrices are identical across layers and the degree distributions are not alternating, Sum A, which applies spectral clustering to the average adjacency matrix, outperforms mFROST for $L<10$, after which the performance difference is marginal. Among the considered methods, only mFROST and DC-MASE consistently improve their performance with the number of layers $L$ until they achieve perfect clustering. mFROST consistently outperforms DC-MASE, with the performance gap being particularly pronounced when the connectivity matrices are identical across layers. For instance, under the same connectivity matrices and different degree corrections, mFROST achieves almost perfect clustering (error $<10^{-3}$) at $L=15$, compared with $L=40$ for DC-MASE. 

The second simulation evaluates the impact of network sparsity on the performance of the methods. We follow the experimental setup of~\cite{Joint_spectral}, using the same configuration as in the first simulation, but fixing the number of layers to $L=20$ and varying the expected edge density of each layer, $\frac{1}{n(n-1)}\sum_{i,j}\mathbb{E}[A_l(i,j)]$, from $2/150$ to $24/150$.    
\begin{figure*}[!t]
    \centering
    \includegraphics[width=0.8\linewidth]{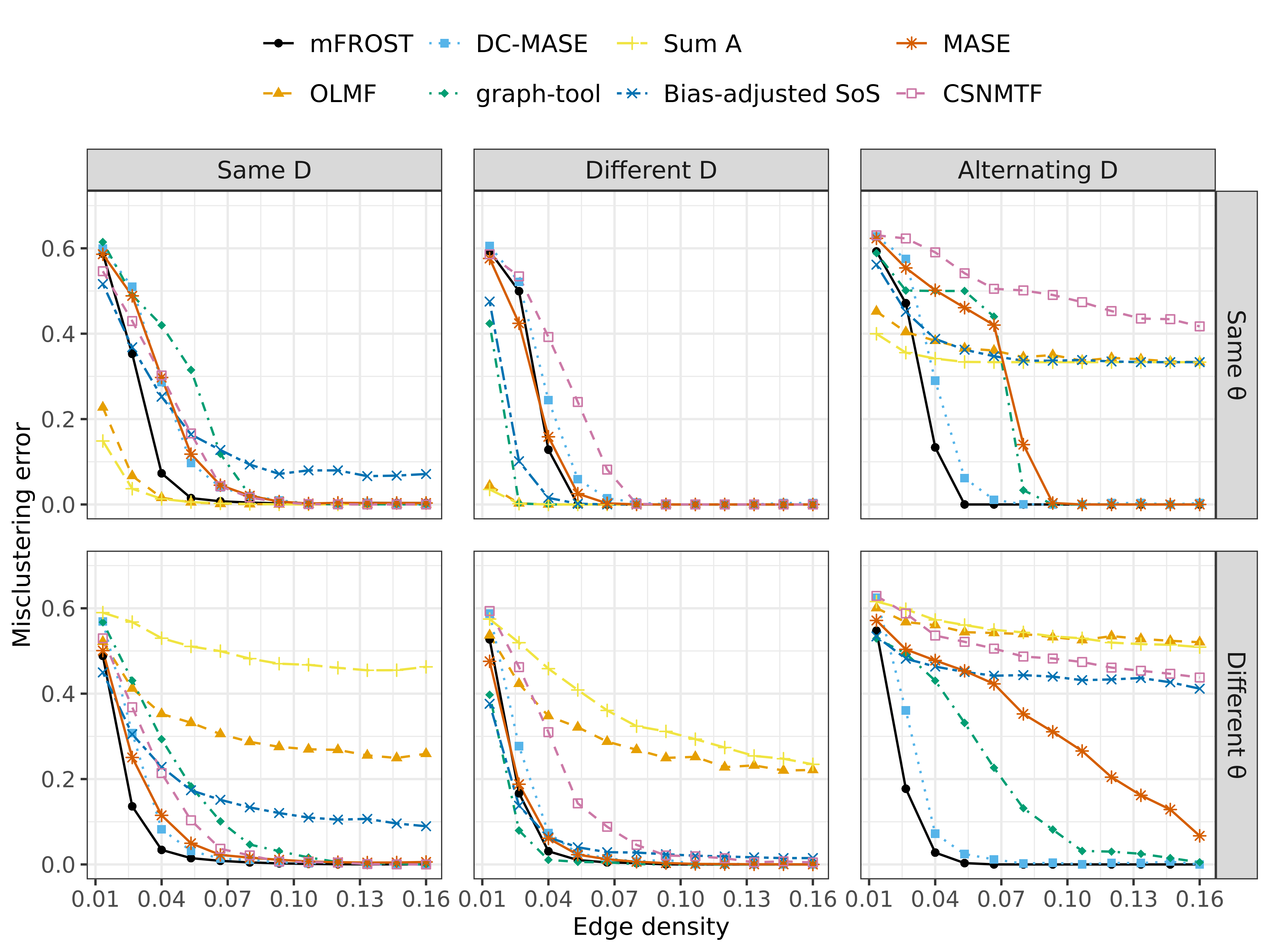}
    \caption{Average misclustering error as a function of the edge density of the networks under different configurations of the connectivity matrices $\theta_l$ and node degree parameters $D_l$. The number of layers is set to $L=20$. }
    \label{fig:sparsityexp}
\end{figure*}
Figure~\ref{fig:sparsityexp} presents the average misclustering error over 100 repetitions as a function of the edge density for each scenario. mFROST, DC-MASE, and graph-tool consistently improve their performance as the edge density increases. mFROST outperforms both DC-MASE and graph-tool in all configurations except when the degree distributions differ across layers. In this latter setting, graph-tool performs slightly better than mFROST: the difference is more pronounced in the very sparse regime when the connectivity matrices are identical across layers, while it becomes marginal when the connectivity matrices differ across layers.

\begin{table*}[t]
\centering
\setlength{\tabcolsep}{3pt}
\caption{Characteristics of the multilayer networks. For layer-dependent statistics, values are reported as mean $\pm$ standard deviation across layers.}
\label{tab:dataset_properties}
\begin{tabular}{lrrrrrrrr}
\hline
\textbf{Network} &
\textbf{$n$} &
\textbf{$L$} &
\textbf{$r$} &
\textbf{Min./max. community size}&
\textbf{Edges} &
\textbf{Density (\%)} &
\textbf{Avg. degree} &
\textbf{Isolated nodes (\%)} \\
\hline
AUCS      & 61   & 5  & 8 & $1/14$   & 620      & $6.8\pm4.0$ & $4.1\pm2.4$    & $26.7\pm26.2$ \\
HS11 & 126 & 4 & 4 &  $ 8/45$ & 2797 & $8.9\pm1.6$& $11.1\pm2.0$& $6.5\pm 2.9$\\
Cora      & 1662 & 2  & 3 & $418/818$   & 27359    & $1.0\pm1.1$ & $16.5\pm18.4$  & $1.1\pm1.6$ \\
Caltech   & 9144 & 6 & 102& $31/800$ & 2621175  & $1.1\pm0.0$ & $95.6\pm2.1$   & $0$ \\
\hline
\end{tabular}

\end{table*}

\subsection{Real-world datasets}

We now evaluate the methods on two real multilayer networks and on two multi-view clustering datasets. For real multilayer networks, ground-truth communities are not always available, and metadata-based labels may not reflect the observed community structure. Multi-view datasets provide class labels that are often associated with similarities between data points and are therefore expected to exhibit assortative structures, with stronger connectivity within communities. Real multilayer networks may exhibit more complex structures, making them useful benchmarks for assessing whether the methods recover meaningful groups under less controlled conditions. We consider the following datasets: 

\paragraph{AUCS~\cite{AUCS}} This multilayer network is composed of five layers representing online and offline relationships (Facebook, leisure, work, co-authorship, and lunch) among 61 employees from the Computer Science department at Aarhus University. The research groups are used as ground truth. 

 \paragraph{HS11~\cite{datasetH11}} This temporal multilayer network records face-to-face contacts among students and teachers at a high school in Marseille, France, over four consecutive days in December 2011. Each layer represents one day, with an undirected edge connecting two individuals if at least one contact was recorded between them that day. Ground-truth communities correspond to the students' school classes, with teachers forming a separate group. The data are available from the SocioPatterns website\footnote{\url{https://www.sociopatterns.org/}}.

\paragraph{The Cora dataset~\cite{Cora}} Following~\cite{NMF_aggreg}, we use a subset of $n=1,662$ machine learning papers
classified into three research areas: Genetic Algorithms, Neural
Networks, and Probabilistic Methods. These labels are used as
ground-truth communities. We construct two layers: a citation layer representing the citation relations between papers and a similarity layer obtained from the cosine similarity between each pair of papers. Each paper is represented by a binary vector indicating the words contained in the paper. The similarity graph is then built by connecting each paper to its 20 nearest neighbors according to cosine similarity. Unlike~\cite{NMF_aggreg}, we do not weight the edges and treat both layers as undirected graphs, to remain consistent with the setting used throughout this paper. Note that mFROST also supports weighted graphs, but we leave this direction for future work.

\paragraph{The Caltech dataset~\cite{li2022caltech}} This dataset contains images of objects from 102 categories, with approximately 30 to 800 images per category, for a total of 9,144 images. For each image, six feature representations are extracted, giving rise to a six-layer network. Each layer is constructed as a k-nearest-neighbor (k-NN) graph, connecting each image to its 50 nearest neighbors based on Euclidean distance as in~\cite{ortiz2024community}.

The main characteristics of each multilayer network used in our experiments are reported in Table~\ref{tab:dataset_properties}. The networks cover a wide range of sizes, numbers of layers, and community structures, with community sizes ranging from relatively balanced to highly unbalanced. AUCS is relatively small with a high proportion of isolated nodes, whereas Caltech is substantially larger, with $2.6$ million edges and 102 communities.

To compare the community assignments produced by the methods with the ground truth, we use normalized mutual information (NMI) and the adjusted Rand index (ARI). NMI quantifies the information shared by the two partitions, normalized by the arithmetic mean of their entropies. ARI measures agreement in whether pairs of nodes belong to the same community, while adjusting for chance. For both metrics, higher values indicate greater agreement, with a value of 1 corresponding to identical partitions.

To account for randomness, we perform 10 independent runs of each method and report the mean and standard deviation of NMI and ARI in Table~\ref{tab:real_networks_results}. mFROST consistently achieves
strong results, obtaining the highest NMI and ARI on HS11, the highest
ARI on Caltech, and second place on both metrics on AUCS. Compared
with OLMF and CSNMTF, which share the same numerical node factor
across layers, mFROST achieves stronger results on most datasets.
Although OLMF achieves a higher NMI and a comparable ARI on Cora,
both OLMF and CSNMTF obtain substantially lower scores on some
other networks; the OLMF implementation also runs out of memory
on Caltech. Graph-tool, which explicitly incorporates degree
correction, is the strongest competitor overall, particularly
on Cora. These results support the competitiveness of mFROST
and complement the synthetic experiments highlighting the benefits
of allowing node sociability to vary across layers.

\begin{table*}[t]
    \centering
    \caption{Mean $\pm$ standard deviation of NMI and ARI
    (both in \%) on real-world multilayer networks.
    Bold and underlined means indicate the highest and
    second-highest scores in each column, respectively. OOM: out of memory.}
    \label{tab:real_networks_results}

    \small
    \setlength{\tabcolsep}{7pt}
    \renewcommand{\arraystretch}{1.15}

    \begin{tabular}{lcccccccc}
    \hline
    \multirow{2}{*}{\textbf{Method}} &
    \multicolumn{2}{c}{\textbf{AUCS}} &
    \multicolumn{2}{c}{\textbf{HS11}}&
    \multicolumn{2}{c}{\textbf{Cora}}&
    \multicolumn{2}{c}{\textbf{Caltech}}\\
    \cline{2-9}
    & \textbf{NMI} & \textbf{ARI}
    & \textbf{NMI} & \textbf{ARI}
    & \textbf{NMI} & \textbf{ARI}
    & \textbf{NMI} & \textbf{ARI} \\
    \hline

    DC-MASE
    & $76.5 \pm 0.6$ & $61.5 \pm 0.4$
    & $68.6 \pm 0.0$ & $75.4 \pm 0.0$ 
    & $38.5 \pm 0.0$ & $32.7 \pm 0.0$
    & $32.0 \pm 0.2$ & $9.6 \pm 0.3$ \\

    Sum A
    & $80.4 \pm 0.0$ & $61.8 \pm 0.0$
    & $78.8 \pm 0.0$ & $83.0 \pm 0.0$ 
    & $44.6 \pm 0.0$ & $37.0 \pm 0.0$
    & $35.3 \pm 0.3$ & $9.0 \pm 0.2$ \\

    Bias-adjusted SoS
    & $76.3 \pm 0.0$ & $59.2 \pm 0.0$
    & $78.6 \pm 0.0$ & $83.0 \pm 0.0$ 
    & $39.2 \pm 0.0$ & $31.2 \pm 0.0$
    & $33.0 \pm 0.0$ & $7.6 \pm 0.1$ \\

    MASE
    & $77.7 \pm 0.0$ & $63.2 \pm 0.0$
    & $80.1 \pm 0.0$ & $\underline{83.0} \pm 0.0$
    & $44.9 \pm 0.0$ & $38.3 \pm 0.0$
    & $32.4 \pm 0.0$ & $9.5 \pm 0.1$ \\

    OLMF
    & $53.5 \pm 0.0$ & $18.1 \pm 0.0$
    & $78.1 \pm 0.0$ & $82.4 \pm 0.0$ 
    & $\underline{53.5} \pm 0.0$ & $\underline{43.5} \pm 0.0$
    & OOM & OOM \\

    graph-tool
    & $\mathbf{84.1} \pm 1.0$ & $\mathbf{69.4} \pm 2.7$
    & $\underline{80.6} \pm 0.0$ & $77.8 \pm 0.0$ 
     & $\mathbf{58.2} \pm 0.3$ & $\mathbf{60.3} \pm 0.4$
    & $\mathbf{39.1} \pm 0.1$ & $\underline{10.8} \pm 0.6$ \\

    CSNMTF
    & $77.4 \pm 4.5$ & $63.9 \pm 6.5$
    & $79.6 \pm 2.6$ & $80.2 \pm 4.0$ 
    & $12.8 \pm 11.0$ & $9.8 \pm 9.9$
    & $19.4 \pm 2.1$ & $5.3 \pm 2.5$ \\

    mFROST
    & $\underline{81.5} \pm 3.7$ & $\underline{67.9} \pm 6.0$
    & $\mathbf{81.5} \pm 0.0$ & $\mathbf{84.9} \pm 0.0$ 
     & $44.9 \pm 1.3$ & $\underline{43.5} \pm 2.7$
    & $\underline{35.4} \pm 0.3$ & $\mathbf{13.1} \pm 0.7$ \\

    \hline
    \end{tabular}
\end{table*}

\subsection{Ablation study}
\label{sec:ablation}
In this section, we investigate the impact of using a shared $Z$~\cite{hribar2026solvingsymmetricmultitypeorthogonal} rather than allowing $Z$ to vary across layers as proposed in our model~\eqref{Joint OtrisymNMF}. We also evaluate the use of the regularized normalized Laplacian, as in OLMF, to reduce the influence of degree heterogeneity. To this end, we implement a variant of mFROST, denoted by mFROST-Zshared, in which the matrix $Z$ is shared across all layers. The only difference lies in the update of $Z$: instead of solving $L$ separate optimization problems, one for each layer, we solve a single optimization problem that minimizes the total reconstruction error across all $L$ layers. We use the same initialization procedure as for mFROST, except that the nonzero entries of the shared $Z$ are initialized using each node’s average degree across layers.

\begin{figure}[!t]
    \centering
    \includegraphics[width=0.96\linewidth]{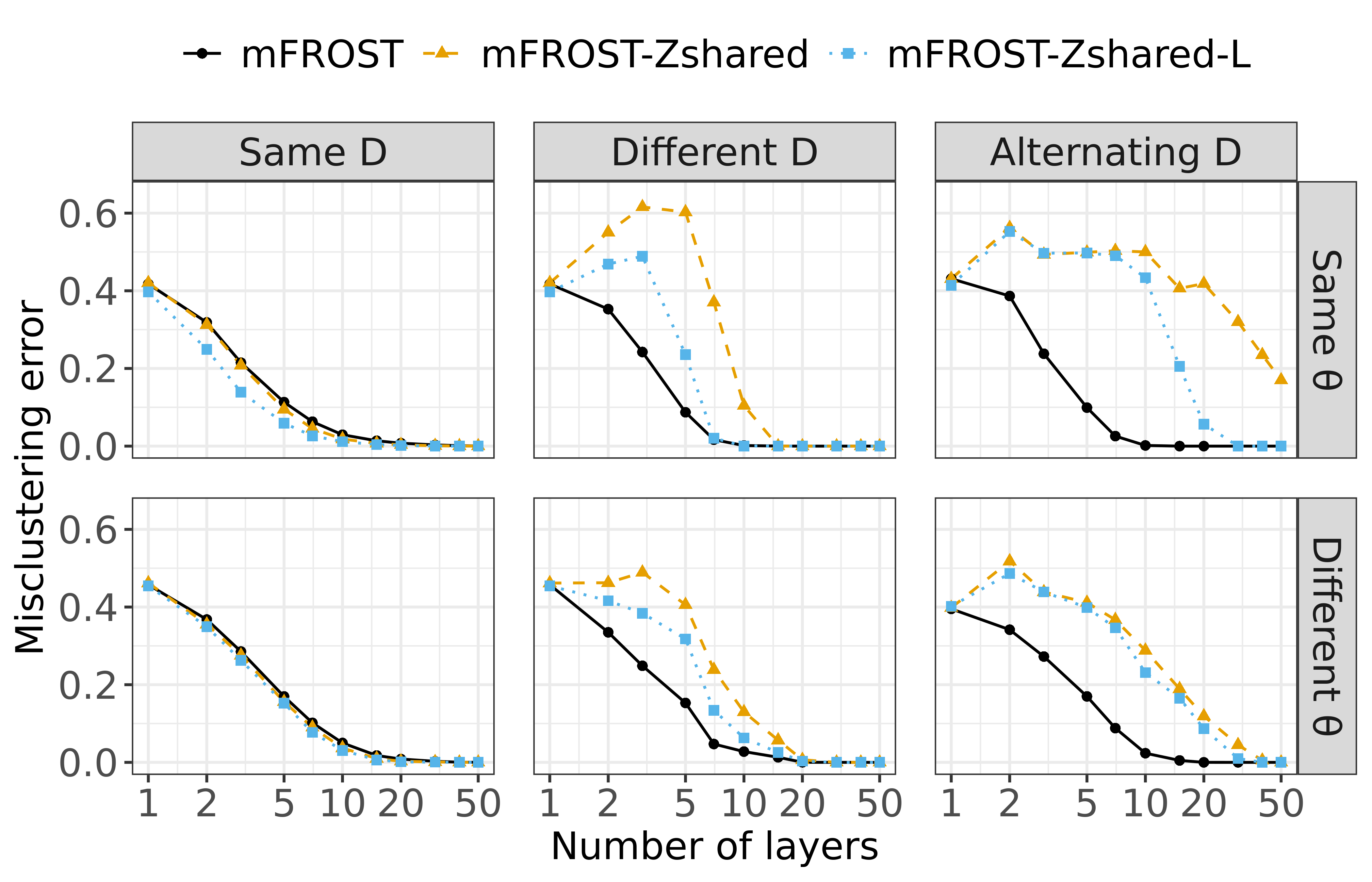}
    \caption{Average misclustering error as a function of the number of layers under different configurations of the connectivity matrices $\theta_l$ and degree-correction parameters $D_l$. }
    \label{fig:sharedZ}
\end{figure}

Figure~\ref{fig:sharedZ} reports the results obtained with mFROST and mFROST-Zshared using either the original adjacency matrices or the regularized graph Laplacian, following the setting of our first synthetic experiment. We observe that, in the same-degree scenario, mFROST achieves results similar to mFROST-Zshared. This shows that, even though mFROST has more parameters to learn than its constrained counterpart, and even in a setting that favors the constrained case, mFROST still achieves excellent results. Using the regularized Laplacian slightly improves the results for mFROST-Zshared. In contrast, when node degrees differ across layers, mFROST clearly outperforms both mFROST-Zshared variants. Using the regularized Laplacian improves the performance of mFROST-Zshared, but the results remain poor compared with mFROST. This gap narrows as the number of layers increases: as the degree parameters are drawn independently across layers, their average tends to stabilize as $L$ grows, reducing the effective degree heterogeneity seen by the shared $Z$. This explains why mFROST-Zshared manages to handle this heterogeneity increasingly well as $L$ increases.

\begin{figure*}[t!]
\centering

\newcommand{\plotMeanBand}[5]{%
    \addplot[
        name path=upper,
        draw=none,
        forget plot
    ]
    table[
        x=iteration,
        y expr=#3*(\thisrow{#2_mean}+\thisrow{#2_std}),
        col sep=comma
    ]{#1};

    \addplot[
        name path=lower,
        draw=none,
        forget plot
    ]
    table[
        x=iteration,
        y expr=#3*(\thisrow{#2_mean}-\thisrow{#2_std}),
        col sep=comma
    ]{#1};

    \addplot[
        fill=#4!,
        fill opacity=0.3,
        draw=none,
        forget plot
    ]
    fill between[of=upper and lower];

    \addplot[
        thick,
        color=#4,
        #5,
        forget plot
    ]
    table[
        x=iteration,
        y expr=#3*\thisrow{#2_mean},
        col sep=comma
    ]{#1};
}

\newcommand{\plotNetwork}[3]{%
    \plotMeanBand
        {Figures/convergence_initialization_#1.csv}
        {#2}{#3}{blue}{solid}
    \plotMeanBand
        {Figures/convergence_random_#1.csv}
        {#2}{#3}{orange}{dashed}
}

\begin{tikzpicture}
\begin{groupplot}[
    group style={
        group name=convergence,
        group size=4 by 2,
        horizontal sep=0.8cm,
        vertical sep=0.8cm,
        xlabels at=edge bottom,
        ylabels at=edge left
    },
    width=0.27\textwidth,
    height=4.5cm,
    xlabel={Iteration},
    grid=major,
    grid style={dotted},
    tick label style={font=\small},
    label style={font=\small},
    title style={font=\small\bfseries},
    scaled ticks=false,
    legend style={
        font=\small,
        draw=none,
        legend columns=2,
        /tikz/every even column/.append style={column sep=1em}
    }
]


\nextgroupplot[
    title={AUCS},
    ylabel={Relative error},
    legend to name=convergencelegend
]
\plotNetwork{AUCS}{error}{1}

\addlegendimage{thick,blue}
\addlegendentry{Our initialization}
\addlegendimage{thick,orange,dashed}
\addlegendentry{Random}

\nextgroupplot[title={HS11}]
\plotNetwork{HS11}{error}{1}

\nextgroupplot[title={Cora}]
\plotNetwork{Cora}{error}{1}

\nextgroupplot[title={Caltech}]
\plotNetwork{Caltech}{error}{1}


\nextgroupplot[
    ylabel={NMI (\%)},
    ymin=0,
    ymax=90
]
\plotNetwork{AUCS}{NMI}{100}

\nextgroupplot[ymin=0, ymax=100]
\plotNetwork{HS11}{NMI}{100}

\nextgroupplot[ymin=0, ymax=50]
\plotNetwork{Cora}{NMI}{100}

\nextgroupplot[ymin=0, ymax=40]
\plotNetwork{Caltech}{NMI}{100}

\end{groupplot}

\path
    (convergence c1r2.south west)
    --
    coordinate[midway] (legendcenter)
    (convergence c4r2.south east);

\node[
    anchor=north,
    yshift=-1.1cm
] at (legendcenter) {
    \pgfplotslegendfromname{convergencelegend}
};

\end{tikzpicture}

\caption{
Relative error (top) and NMI (bottom) at each iteration of
mFROST on AUCS, HS11, Cora, and Caltech, using the proposed
and random initialization. Solid and dashed lines show the
mean values over 10 restarts, respectively, while shaded
areas represent one standard deviation.
}
\label{fig:convergence}
\end{figure*}
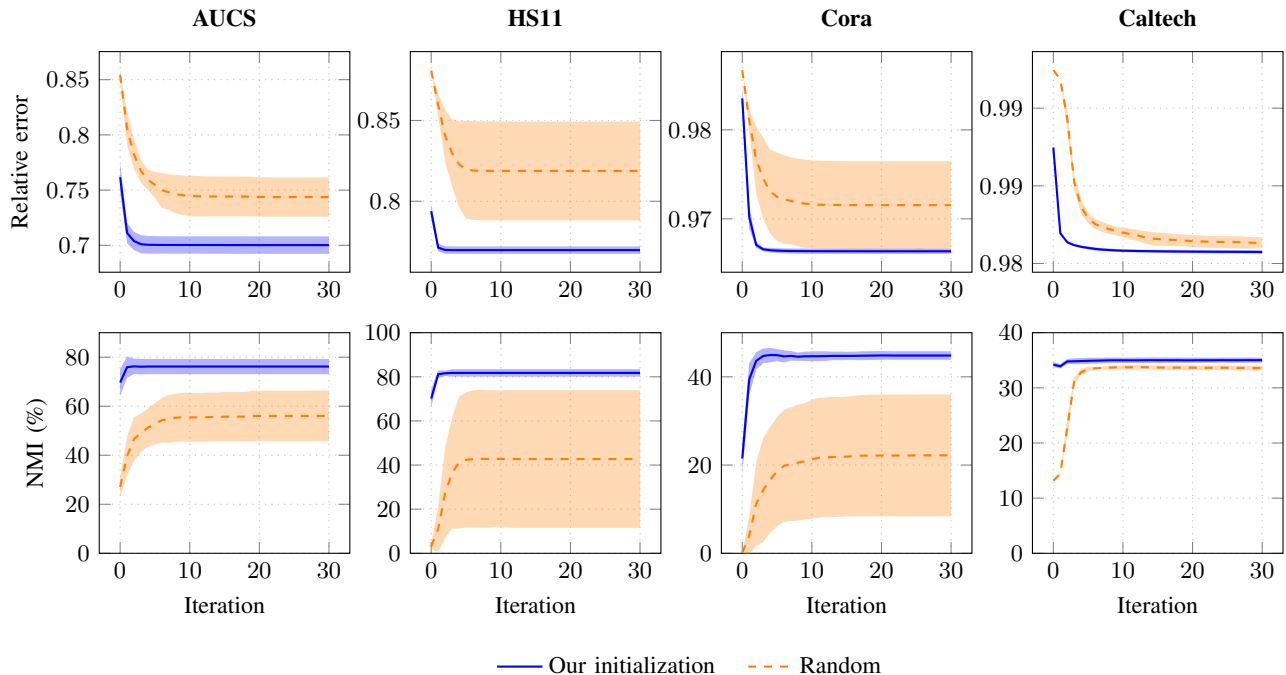
\subsection{ Empirical stability and convergence}
\label{Section: scalability}
We additionally investigate the empirical stability and convergence of mFROST. Our goal is to assess whether the computational cost of mFROST remains practical as the size of the multilayer network increases. Each iteration of mFROST has a computational cost of $
O\left(r\sum_{\ell=1}^{L}\operatorname{nnz}(A_\ell)\right) $. This is of the same order as the cost of sparse matrix operations used in spectral methods. Since mFROST is iterative, the overall computational cost also depends on the number of iterations required to reach the stopping criterion. The algorithm stops when the difference in relative error between two consecutive iterations falls below $\delta=10^{-5}$. 

Table~\ref{tab:runtime} reports the mean and standard deviation of runtime, NMI, and number of iterations over 10 individual restarts of mFROST on each real-world dataset. Across all experiments, mFROST requires only a few iterations, the stopping criterion is reached within at most 20 iterations. Regarding runtime, the observed times are consistent with a complexity that scales approximately with the number of communities times the number of edges, since the number of iterations remains similar across networks. For Caltech with 2.6 million edges and 102 communities, a single restart with $\delta=10^{-5}$ takes about 4 minutes. For each network, the mean NMI across individual restarts is close to the mean NMI reported in Table~\ref{tab:real_networks_results}, where each run retains the restart with the lowest reconstruction error. The standard deviation of NMI remains small across restarts (less than 5\%), showing that mFROST reliably reaches solutions of similar quality although our initialization is randomized. This stability stems from the use of the smoothed separable NMF method in our initialization procedure, which provides a data-driven starting point rather than a purely random one, consistent with our previous findings in the single-layer case~\cite{dache2026matrixfactorizationframeworkcommunity}.

\begin{table}[t]
\centering
\caption{ Mean $\pm$ standard deviation of runtime, NMI, and number of iterations over 10 restarts of mFROST on real-world datasets.}
\label{tab:runtime}
\begin{tabular}{lccccc}
\hline
\textbf{Network} & \textbf{Edges} & \textbf{$r$} & \textbf{Time (s)} & \textbf{Iterations} & \textbf{NMI (\%)} \\
\hline
AUCS     & 620   & 8 & $0.19\pm0.08$ & $11.0\pm 3.5 $ &$78.0\pm4.8$ \\
HS11 & 2797& 4& $0.17\pm0.05$ &$5.3\pm1.0$ & $81.2\pm0.6$ \\
Cora     & 27359 & 3 &$1.07\pm0.28$ &$10.9\pm3.4$ &$44.1\pm 1.3$ \\
Caltech  & 2621175 & 102 & $232\pm43$ &$13.7\pm 2.4$ &$35.2\pm0.3$ \\
\hline
\end{tabular}

\end{table}

We further examine the empirical convergence of mFROST and compare our initialization strategy with random initialization. Figure~\ref{fig:convergence} shows the relative reconstruction error and NMI as functions of the iteration number for each multilayer network, averaged over 10 restarts for each initialization strategy. Both quantities stabilize after relatively few iterations. Our initialization generally leads to faster convergence, a higher mean NMI, and lower variability across restarts than random initialization. We also observe that, on average, NMI generally increases as the reconstruction error decreases, suggesting that improvements in reconstruction quality tend to coincide with improvements in community recovery. The stopping criterion is based solely on changes in reconstruction error, NMI is reported only for evaluation.

\section{Conclusion}
\label{sec:conclusion}
In this paper, we proposed mFROST, a community detection algorithm for multilayer networks based on a new joint nonnegative symmetric matrix trifactorization model. Our model recovers a common community structure while allowing community interactions and node sociability to vary across layers. Its close relationship with the multilayer degree-corrected stochastic block model (MDCBM) provides an interpretable representation of community memberships, degree heterogeneity, and interactions between communities.

A key novelty of our joint factorization formulation is that it includes a sociability parameter for each node in each layer as in the MDCBM. Unlike joint factorization approaches that impose the same node factors across layers, our model allows the relative sociability of nodes within each community to vary across layers while preserving common community assignments. This flexibility enables mFROST to account for degree heterogeneity both within and across layers. 

Experiments on synthetic and real-world multilayer networks show the consistently strong performance of mFROST across diverse community interactions and degree distributions, including settings with isolated nodes. Several state-of-the-art methods achieve good results in specific configurations but suffer substantial losses in accuracy in others. mFROST remains competitive across these settings, highlighting its ability to accommodate structural heterogeneity within and across layers. The ablation study shows that explicitly modeling layer-specific node sociabilities improves performance when node degrees vary across layers, without compromising performance when they remain the same across layers. The proposed block coordinate descent algorithm, mFROST, yields a non-increasing objective and exhibits rapid empirical convergence, with a per-iteration cost that scales linearly with the product of the number of edges and the number of communities. Our initialization strategy further improves convergence speed and reduces variability in clustering quality across restarts.

Future work could extend mFROST to partially observed multilayer networks, where some interactions or nodes are unobserved in individual layers. Such settings arise in social and biological networks with incomplete observations~\cite{missing_nodes}. A natural extension would restrict the reconstruction error to observed entries. The shared community assignments would allow information from other layers to contribute to the clustering of partially observed nodes. Another direction is to allow overlapping communities through mixed memberships, following the degree-corrected mixed-membership framework~\cite{MMDCBM}. This extension would require replacing the orthogonality constraints, which currently enforce disjoint communities.

\section*{Acknowledgment}
This research is funded by the European Union (ERC consolidator, eLinoR, no. 101085607). Alexandra Dache is a Research Fellow of the Fonds de la Recherche Scientifique - FNRS (F.R.S.-FNRS).


\bibliographystyle{IEEEtran}
\bibliography{refs}


 





\end{document}